\documentclass[fontsize=11pt, paper=letter, parskip=half]{scrartcl}

\usepackage[margin=1.25in]{geometry}
\usepackage[T1]{fontenc}
\usepackage[utf8]{inputenc}
\usepackage{amsmath}
\usepackage{graphicx}
\usepackage{longtable,booktabs,array}
\usepackage{calc}
\usepackage{float}
\usepackage{url}
\usepackage[colorlinks=true,linkcolor=black,citecolor=blue,urlcolor=blue]{hyperref}

\newcommand{\citep}[1]{\cite{#1}}
\newcommand{\citet}[1]{\cite{#1}}
\newcommand{\biburl}[1]{\href{#1}{\textit{Available online}}}
\providecommand{\tightlist}{%
  \setlength{\itemsep}{0pt}\setlength{\parskip}{0pt}}

\title{The Reaction of Financial Markets to the Releases of DeepSeek R1 and Kimi K3: A Critical Analysis}
\author{Marco Taboga\thanks{The opinions expressed in this paper are the
author's and do not necessarily reflect those of Banca d'Italia. I thank Marco Albori, Claudia Biancotti, Fabrizio Ferriani and Giovanni Veronese for providing feedback on the first draft of the paper.}\\
Banca d'Italia}
\date{\today}

\begin{document}
\pagestyle{empty}
\maketitle
\thispagestyle{empty}

\begin{abstract}
This paper compares financial-market reactions to the releases of
DeepSeek R1 in January 2025 and Kimi K3 in July 2026. Both AI models were
presented as credible Chinese competitors near the capability frontier,
but their releases conveyed different economic information. DeepSeek R1 combined
open weights with unusually low reported training and inference costs,
prompting investors to reassess expected demand for computing
infrastructure and producing large stock-price decreases among semiconductor
and data-centre-related firms. Kimi K3 instead entered at broadly
frontier prices and generated a market reaction concentrated more heavily
among rival Chinese AI laboratories. Drawing on descriptive event-window
evidence, model capability and cost comparisons, subsequent industry
developments, and the historical diffusion of open-source software, the
paper argues that open-weight models redistribute rents across the AI
value chain rather than eliminating them. Incumbents remain protected by
complementary assets including distribution, product bundling, switching
costs, proprietary data, serving capacity, and contractual trust. The
analysis suggests that investors and policy makers should assess risks looking not only at model
capability, but also at the ownership and durability of complementary assets
across the AI value chain.
\end{abstract}



\newpage
\tableofcontents
\newpage

\section{Introduction}\label{introduction}

In January 2025, the release of a large language model by a Chinese
hedge-fund affiliate became a macro-financial event: DeepSeek R1 was an
open-weight\footnote{An open-weight model makes its trained numerical
  parameters (the so-called weights) publicly available, allowing users to run and often
  fine-tune it on their own infrastructure. The term does not necessarily
  imply that the training data, source code, or full development process
  is also public.} reasoning model, distributed under a permissive licence and
served through an API at prices far below those of frontier models. Two
claims travelled with it: that it approached the capabilities of the
best Western closed models, and that it had been trained at a small
fraction of their cost. Equity markets read the second claim as news
about the demand for compute and repriced the entire chain that supplies
it---AI-chip vendors, semiconductor fabs, and companies financing data
centres and AI labs.

Eighteen months later, in July 2026, Moonshot AI, another Chinese AI
lab, released Kimi K3, a very large model positioned by its
capabilities, benchmark scores, and pricing as a frontier product rather
than a discount substitute. The reaction of equity markets was almost
equally strong, although with differences that are informative. Drops in broad-market and technology-sector equity indices were of a similar order of magnitude, but the cross-section
of returns was quite different: the sharpest losses were recorded not by US
chipmakers or US model developers, but by other Chinese AI labs.

This paper uses the two episodes as comparative case studies to answer
three related questions. First, what economic information did each model
release convey about the cost and production of frontier AI? Second, why
did the two releases coincide with similar aggregate equity-market
declines but markedly different patterns of losses across firms? Third,
which parts of the AI value chain are most exposed when a Chinese model
approaches the capability frontier, and which complementary assets allow
incumbents to preserve their market position? By combining model-level
evidence on capability, price, and inference efficiency with descriptive
event-window returns, historical comparisons, and subsequent industry
developments, the paper evaluates the hypothesis that open-weight
competition redistributes rents across the AI value chain rather than
eliminating them in aggregate.

Section~\ref{description-of-the-two-models} compares the capabilities, architecture, API prices, and
task-level costs of the two models with those of their contemporary
closed competitors. It finds that both releases were near the capability
frontier, but conveyed very different cost signals. DeepSeek R1 offered
reasoning performance at prices roughly an order of magnitude below the
closed frontier, whereas Kimi K3 was positioned and priced as a frontier
product. Once its relatively high token consumption is taken into
account, Kimi K3's cost per completed task was only modestly below that
of the leading US closed model and did not represent a comparable cost
discontinuity.

Section~\ref{reaction-of-financial-markets-to-the-two-releases} compares the timing, magnitude, and cross-sectional composition
of the two market reactions. Although the main broad-market and technology-sector equity indices recorded similar declines in the two episodes, the January 2025 losses were concentrated in the
compute supply chain, while the largest reported losses in July 2026
were associated with rival Chinese AI laboratories. The section argues
that DeepSeek R1 was interpreted primarily as news about the production
function for AI and future demand for advanced chips, whereas Kimi K3 was more
plausibly news about the arrival of an additional credible supplier at
the existing frontier price point. Because both windows contain major
confounding events, these patterns are treated as descriptive rather
than causal evidence.

Section~\ref{who-is-most-exposed-to-chinese-competition} develops the hypothesis suggested by this cross-section: the
firms most directly exposed to Kimi K3 may be other laboratories selling
near-frontier models without the complementary assets held by large US
incumbents. It shows that customers often purchase an assembled service rather
than model weights alone. Product bundling, installed-user switching
costs, differentiated applications, serving infrastructure, contractual
trust, and proprietary interaction data can therefore protect incumbent
rents even when comparable weights become openly available. The section
also notes that the durability of the proprietary-data advantage is
uncertain, since it depends on the still-unresolved effectiveness and
cost of model distillation.

Section~\ref{a-historical-precedent} places the arguments above in the context of the diffusion of
open-source software in the Nineties. That history suggests that a free substitute for
one layer of a technology stack need not eliminate profits throughout
the industry: it can expand demand, segment the market, and shift value
towards infrastructure, integration, support, distribution, and the
customer relationship. The analogy therefore predicts a redistribution
of AI rents rather than the displacement of all closed-model providers,
while also warning that firms concentrated in the commoditised layer can
still suffer substantial losses.

Section~\ref{ai-industry-developments-after-the-two-releases} evaluates these mechanisms against industry developments after
each release. In the eighteen months following DeepSeek R1, open-weight
and Chinese-model usage rose sharply, but closed-model usage also
continued to grow and demand for accelerators did not decline. US
laboratories responded in part by releasing open-weight models of their
own. These developments suggest that
lower model costs expanded the volume of AI use and redistributed demand
within the industry rather than reducing the value of demand for the technology stack
as a whole. The immediate aftermath of Kimi K3 revealed a different set of
constraints and competitive responses. Moonshot AI temporarily paused
new subscriptions when demand exceeded its serving capacity,
demonstrating that frontier-class weights alone do not constitute a
scalable service. US incumbents responded with rapid product releases.

Section~\ref{conclusions} concludes that the competitive threat is real
but unevenly distributed: firms with durable complementary assets are
better protected, while those dependent on the commoditising model layer
remain exposed.

\section{Description of the Two
Models}\label{description-of-the-two-models}

Both releases shared one feature that is essential for interpreting the
market reaction: at the moment of release, each model was close to the
capability frontier, rather than a lagging imitation of it.

DeepSeek R1 was a large AI model (671 billion total parameters) having relatively parsimonious computational requirements during utilization (only 37 billion parameters were activated per inference pass). It was published on 20 January 2025 under a licence allowing free reproduction and broad reuse. Unusually for the time, it showed the reasoning leading to its answers. 

\begin{figure}[H]
\centering
\caption{DeepSeek R1 versus frontier models at release}
\label{fig-deepseek-benchmark-comparison}
\vspace{0.3em}
\includegraphics[width=\textwidth]{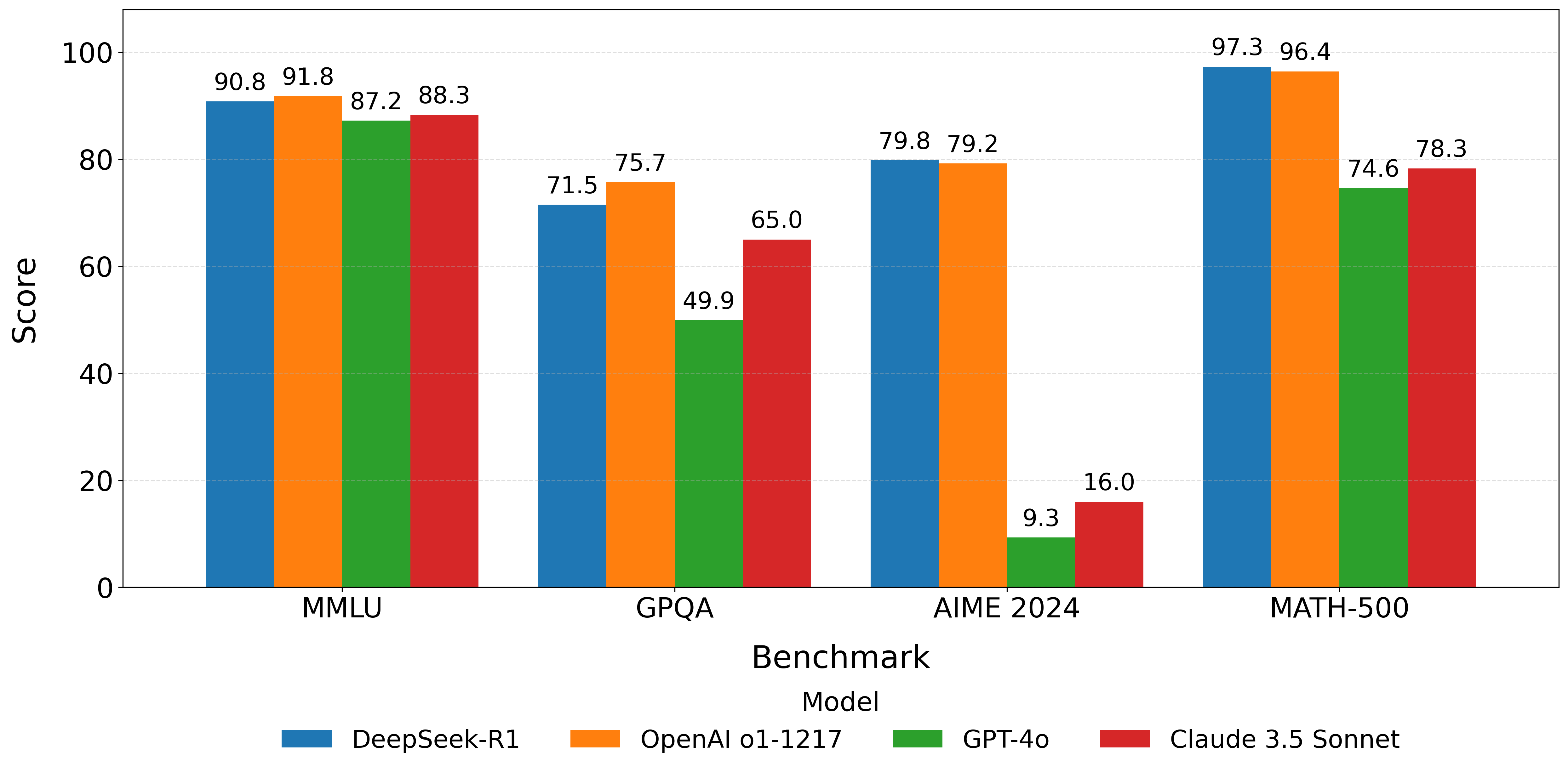}
\par\vspace{0.3em}
\parbox{\textwidth}{\footnotesize\textit{Note:} MMLU measures broad academic knowledge and reasoning across 57 subjects; GPQA comprises expert-written, graduate-level questions in biology, physics, and chemistry; AIME 2024 uses problems from the 2024 American Invitational Mathematics Examination; and MATH-500 is a 500-problem subset of the MATH benchmark for competition-level mathematical reasoning.}
\end{figure}

Figure \ref{fig-deepseek-benchmark-comparison} reports the
benchmark scores published at release for DeepSeek R1 alongside the
contemporary closed frontier. On mathematical and scientific reasoning
tasks, R1 was broadly at parity with OpenAI's o1 and clearly ahead of
the general-purpose models then in wide commercial use.

\begin{table}[H]
\centering
\caption{API prices of DeepSeek R1 and contemporary frontier models}
\label{tbl-r1-frontier-prices}
\vspace{0.3em}
\small
\renewcommand{\arraystretch}{1.25}
\begin{tabular}{@{}>{\raggedright\arraybackslash}p{0.48\textwidth}
                    >{\centering\arraybackslash}p{0.20\textwidth}
                    >{\centering\arraybackslash}p{0.20\textwidth}@{}}
\toprule
\textbf{Model} & \textbf{Input price} & \textbf{Output price} \\
\midrule
DeepSeek R1 & \$0.55 & \$2.19 \\
OpenAI o1-preview & \$15.00 & \$60.00 \\
GPT-4o (Nov. update) & \$2.50 & \$10.00 \\
Claude 3.5 Sonnet (new) & \$3.00 & \$15.00 \\
\bottomrule
\end{tabular}
\par\vspace{0.5em}
\parbox{\dimexpr0.88\textwidth+4\tabcolsep\relax}{\footnotesize\textit{Note:} Prices are in US dollars per million tokens.}
\end{table}

What distinguished R1, however, was price rather than capability. As
Table \ref{tbl-r1-frontier-prices} shows, its API price was more than an
order of magnitude below that of the closest competitor on
reasoning tasks, and well below that of general-purpose frontier models.
Since the weights were openly distributed, the price of serving the
model was in addition bounded by the cost of renting accelerators,
rather than set by a single vendor.

Technical commentary attributed the cost-effectiveness of R1 to its
architecture, to a training pipeline leaning heavily on reinforcement
learning, and to engineering ingenuity under the constraint of United
States export controls on accelerators \citep{mercer2025}. A later
peer-reviewed account put the compute cost of R1's reinforcement-learning
stage (one of the latest stages of training) at approximately US\$294,000 \citep{nature2025}. The cost of training the base model (the first stages of training) was estimated to be US\$5.576 million,
excluding prior research and ablation experiments
\citep{deepseekv32024}. 
These sums were far below widely cited estimates for other 
frontier training runs completed around that time: the Stanford AI Index estimated compute costs of
approximately US\$79 million for GPT-4, US\$192 million for Gemini 1.0
Ultra, and US\$170 million for Llama 3.1-405B
\citep{aiindex2025}. It is important to note that DeepSeek's cost effectiveness, to which markets reacted, was mostly self-reported and subsequent analyses raised doubts about DeepSeek's claims \citep{patel2025deepseek}. 

\begin{figure}[H]
\centering
\caption{Kimi K3 versus frontier models at release}
\label{fig-kimi-benchmark-comparison}
\vspace{0.3em}
\includegraphics[width=\textwidth]{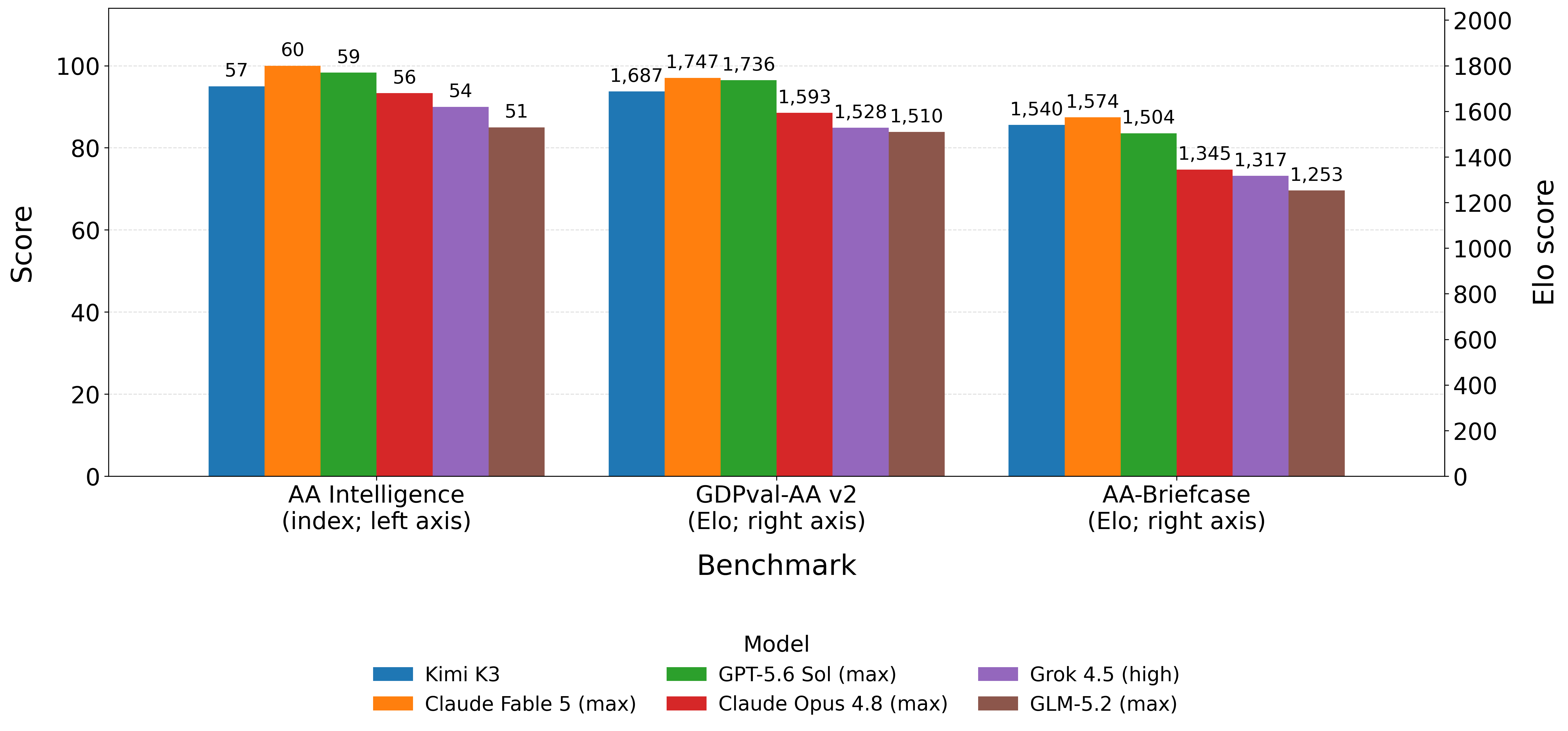}
\par\vspace{0.3em}
\parbox{\textwidth}{\footnotesize\textit{Note:} The Artificial Analysis (AA) Intelligence Index is a composite measure of model performance across reasoning, coding, scientific knowledge, and related capabilities. GDPval-AA v2 evaluates agentic performance on economically valuable real-world tasks spanning a broad range of occupations and industries. AA-Briefcase evaluates long-horizon agentic knowledge work through realistic business projects requiring deliverables such as spreadsheets, presentations, and memoranda.}
\end{figure}

Kimi K3 was much larger than R1: it was a
2.8-trillion-parameter model with a one-million-token context window,
positioned by Moonshot AI as an open model with weights
scheduled for public release \cite{artificialanalysis2026,moonshot2026}.

On the Artificial Analysis Intelligence Index---a composite of
evaluations spanning reasoning, coding, and knowledge---it scored 57 at
release, ranking third overall and comparable to Opus 4.8,
but behind Fable 5 and GPT-5.6 Sol
\cite{artificialanalysis2026}. Figure \ref{fig-kimi-benchmark-comparison} places these evaluations alongside the performances on two agentic
benchmarks, which are informative measures for commercial use
because they score the completion of multi-step tasks rather than the
answering of isolated questions. On both, Kimi K3 was very near
the frontier at release: it ranked above Opus 4.8 on
GDPval-AA v2, and was second only to Claude Fable 5 on AA-Briefcase, an
evaluation of agentic knowledge work.

In short, the performance of Kimi K3 on popular benchmarks was, like that
of DeepSeek R1, close to that of contemporary frontier models. 

\begin{table}[H]
\centering
\caption{API prices, output speed, and Intelligence Index evaluation costs of Kimi K3 and contemporary frontier models}
\label{tbl-frontier-model-prices-speed-cost}
\vspace{0.3em}
\small
\renewcommand{\arraystretch}{1.25}
\begin{tabular}{@{}>{\raggedright\arraybackslash}p{0.36\textwidth}
                    >{\centering\arraybackslash}p{0.13\textwidth}
                    >{\centering\arraybackslash}p{0.13\textwidth}
                    >{\centering\arraybackslash}p{0.13\textwidth}
                    >{\centering\arraybackslash}p{0.13\textwidth}@{}}
\toprule
\textbf{Model} & \textbf{Input \ \  price} & \textbf{Output price} &
\textbf{Speed (tok/s)} & \textbf{Evaluation cost} \\
\midrule
Kimi K3 & \$3.00 & \$15.00 & 32.9 & \$2,437.41 \\
Claude Fable 5 (max) & \$10.00 & \$50.00 & 73.4 & \$5,630.52 \\
GPT-5.6 Sol (max) & \$5.00 & \$30.00 & 68.7 & \$3,442.81 \\
Claude Opus 4.8 (max) & \$5.00 & \$25.00 & 57.6 & \$3,752.55 \\
Grok 4.5 (high) & \$2.00 & \$6.00 & 52.0 & \$639.87 \\
GLM-5.2 (max) & \$1.40 & \$4.40 & 176.8 & \$712.45 \\
\bottomrule
\end{tabular}
\par\vspace{0.5em}
\parbox{\dimexpr0.88\textwidth+8\tabcolsep\relax}{%
\footnotesize\textit{Note:} Input and output prices are in US dollars per
million tokens. Speed is measured in output tokens per second. Evaluation cost
is the total cost in US dollars to run all evaluations in the Artificial
Analysis Intelligence Index.}
\end{table}

Crucially, and in sharp contrast with DeepSeek R1, Kimi K3 was not
priced as a discount substitute. Its published API prices (Table \ref{tbl-frontier-model-prices-speed-cost}) were \$3.00
per million input tokens and \$15.00 per million output tokens, roughly half the cost of GPT-5.6 Sol, but much higher than that of other slightly less performing models such as Grok 4.5 and GLM-5.2.

The price of a token is, however, an incomplete measure of the cost of inference for a
reasoning model, because such models differ greatly in how many tokens
they consume to reach an answer. Relative to the broader set of models
evaluated on the Artificial Analysis benchmark, Kimi K3 was verbose: it
generated approximately 130 million output tokens across the evaluation
suite, against a median of about 63 million for other models
\cite{techtimes2026}.

The evaluation cost reported in Table
\ref{tbl-frontier-model-prices-speed-cost} incorporates this token
consumption and therefore provides a more informative comparison than
the posted token prices alone. Running the full Intelligence Index suite
cost \$2,437.41 with Kimi K3: 29.2\% less than with GPT-5.6 Sol, 35.0\%
less than with Claude Opus 4.8, and 56.7\% less than with Claude Fable 5.
Kimi K3 was nevertheless 3.8 times as expensive to evaluate as Grok 4.5
and 3.4 times as expensive as GLM-5.2. It was therefore relatively
economical among the closest frontier models, but not a low-cost model
within the broader comparison set.

Performance and cost per task are not the only economically relevant
factors. Speed is also important, and Kimi K3 was the slowest model in
the comparison, generating 32.9 output tokens per second. Claude Fable 5
and GPT-5.6 Sol were respectively about 2.2 and 2.1 times faster, Claude
Opus 4.8 and Grok 4.5 were about 1.8 and 1.6 times faster, and GLM-5.2
was about 5.4 times faster. Kimi K3's lower evaluation cost relative to
the closest frontier models therefore came with a substantial
latency disadvantage, while the cheaper Grok and GLM alternatives were
also faster.

Overall, when token efficiency and speed are considered together, Kimi
K3 did not clearly dominate the closed frontier on capability per dollar
and offered no DeepSeek-like cost
discontinuity relative to GPT-5.6 Sol. We conclude that Kimi K3 is best
understood as a near-frontier model priced like one.

\section{Reaction of Financial Markets to the Two
Releases}\label{reaction-of-financial-markets-to-the-two-releases}

The market reaction was strong in both cases. Table
\ref{tbl-release-comparison} compares the two episodes using global and
US broad-market equity indices, sectoral indices, the share prices of major US
technology companies, and available market-based indicators for US and
Chinese AI laboratories.

Some technical remarks about the data displayed in the table are in order.

First, the financial instruments considered are heterogeneous. Most
observations refer to listed securities or equity indices, whereas the
entries for Anthropic and OpenAI refer to pre-IPO futures traded on
cryptocurrency exchanges. The information conveyed by these instruments
may be limited by low liquidity, market design, and the composition of
participants. Although they broaden the available evidence, their
returns are not directly comparable with common-equity returns.

Second, the selection of event windows necessarily involves some
judgement. Market commentary suggests that the reaction to DeepSeek was
delayed: although the model was released on 20 January 2025, prices moved
primarily between 24 and 27 January, after adoption data and third-party
analyses had circulated. By contrast, the reaction to Kimi K3 was almost
immediate. The windows used here follow the timing identified in market
commentary, but their precise boundaries are somewhat arbitrary.

Third, both event windows contain substantial confounding news. Each
coincided with an earnings season, while the July 2026 window also
overlapped with a broader repricing of semiconductor stocks and a
re-escalation of the US--Iran war. These confounding factors may have amplified or
offset the release-related effects; consequently, the observed declines
cannot generally be interpreted even as upper bounds. No attempt is made
to isolate the component attributable specifically to either model
release.

With these technical remarks in mind, the figures in Table \ref{tbl-release-comparison} are therefore to be interpreted with caution, as
unadjusted event-window returns, not estimates of causal effects. 

\begin{table}[H]
\centering
\caption{Comparison of the DeepSeek R1 and Kimi K3 releases and market reactions}
\label{tbl-release-comparison}
\vspace{0.3em}
\small
\renewcommand{\arraystretch}{1.3}
\newcommand{\marketreturn}[1]{\makebox[3.5em][r]{$#1\%$}}
\newcommand{\missingreturn}{\makebox[3.5em][c]{N/A}}
\begin{tabular}{@{}>{\raggedright\arraybackslash}p{0.32\textwidth}
                    >{\raggedright\arraybackslash}p{0.28\textwidth}
                    >{\raggedright\arraybackslash}p{0.32\textwidth}@{}}
\toprule
 & \textbf{DeepSeek R1} & \textbf{Kimi K3} \\
\midrule
Release date & 20 Jan. 2025 & 16 Jul. 2026 \\
Market reaction type & Delayed, after first adoption data and analyses & Immediate \\
Market reaction window & 24--27 Jan. 2025 & 15--17 Jul. 2026 \\
Confounding factors in window & Earnings season & Earnings season\newline Semiconductor repricing\newline Re-escalation of US--Iran war \\
\midrule
\textbf{Global equity indices} &  &  \\
DS World & \marketreturn{-0.9} & \marketreturn{-1.7} \\
DS World Technology & \marketreturn{-4.2} & \marketreturn{-4.4} \\
DS World Semiconductors & \marketreturn{-12.0} & \marketreturn{-6.7} \\
SOX & \marketreturn{-9.1} & \marketreturn{-5.8} \\
 &  &  \\
\textbf{US equity indices} &  &  \\
S\&P 500 & \marketreturn{-1.5} & \marketreturn{-1.5} \\
Nasdaq 100 & \marketreturn{-3.0} & \marketreturn{-3.1} \\
 &  &  \\
\textbf{US equities} &  &  \\
Alphabet & \marketreturn{-4.2} & \marketreturn{-6.5} \\
Amazon & \marketreturn{0.2} & \marketreturn{-3.0} \\
Meta & \marketreturn{1.9} & \marketreturn{-5.2} \\
Microsoft & \marketreturn{-2.1} & \marketreturn{-0.5} \\
NVIDIA & \marketreturn{-17.0} & \marketreturn{-4.6} \\
 &  &  \\
\textbf{US labs (pre-IPO futures)} &  &  \\
Anthropic USDT & \missingreturn & \marketreturn{-1.8} \\
OpenAI USDT & \missingreturn & \marketreturn{-5.0} \\
 &  &  \\
\textbf{Chinese labs} &  &  \\
Alibaba & \marketreturn{2.9} & \marketreturn{-0.7} \\
MiniMax Group & \missingreturn & \marketreturn{-17.1} \\
Tencent & \marketreturn{1.4} & \marketreturn{-1.3} \\
Xiaomi & \marketreturn{0.7} & \marketreturn{3.9} \\
Zhipu AI & \missingreturn & \marketreturn{-35.1} \\
\bottomrule
\end{tabular}
\par\vspace{0.5em}
\parbox{\dimexpr0.92\textwidth+4\tabcolsep\relax}{\footnotesize\textit{Note:} Reported market reactions are percentage price changes between the closing price on the first day of the reaction window and the closing price on the last day of the reaction window. DS (Datastream) indices and other equity prices are sourced from LSEG.}
\end{table}

The aggregate magnitudes of the market reactions to the two releases
were remarkably similar---the S\&P 500 lost 1.5\% in both cases, while
the Nasdaq 100 fell by 3.0\% and 3.1\%, respectively---but their
composition differed substantially. 

In January 2025, losses were
concentrated in the compute supply chain, with NVIDIA down 17.0\% and the
SOX semiconductor index down 9.1\%. In July 2026, the
corresponding declines were smaller, at 4.6\% and 5.8\%, while the
largest losses reported in the table were associated with other Chinese
AI laboratories: Zhipu AI fell 35.1\% and MiniMax Group 17.1\%. Before the release of Kimi K3, Zhipu AI's
GLM 5.2 was the highest-scoring open-weight model on the Artificial
Analysis Index, some five points below the best closed model, and
MiniMax's M3 occupied the adjacent niche as the only open model
accepting native image and video input \cite{clark2026}. 

According to market commentaries \cite{carew2025}, part of the
reaction to DeepSeek was motivated not by the model's capabilities but
by the claims of unusually high training efficiency documented in Section~\ref{description-of-the-two-models}. If a frontier-class reasoning model could be
trained and served for a small fraction of the outlays assumed in
analysts' models, then the projected demand for AI chips, and the
capital expenditure plans predicated on it, were too high. That was news
about the production function for AI, and it propagated naturally to
chipmakers.

The Kimi K3 release carried no comparable claim. The
cross-sectional returns were instead consistent with news about the
number of credible model suppliers at a given price point, with the
largest effects falling on laboratories competing at the same layer
of the value chain without equally strong complementary assets.\footnote{Moonshot AI itself was treated differently: in the days following the
release, it was reported to be pursuing a Hong Kong IPO at a valuation
of approximately \$30 billion, supported by annual recurring revenue of
\$300 million as of June 2026 \cite{techtimes2026}.} Tellingly, Alibaba, Tencent and Xiaomi did not
experience significant stock-market losses after the release, despite
controlling large AI laboratories that regularly publish high-quality
open-weight models (Qwen, Hy and MiMo, respectively). These companies are diversified
conglomerates with distribution, infrastructure, hardware, and customer
relationships through which model capabilities can be monetised. 

The evidence for US laboratories points in the same direction, although it is less straightforward to interpret. In the Kimi K3 window, the pre-IPO futures linked to Anthropic and
OpenAI declined by 1.8\% and 5.0\%, respectively—material moves, but much smaller than
those of the listed Chinese pure-play laboratories. The prices of the large US technology
companies that control their own AI laboratories, notably Alphabet and Meta, also fell,
but by far less than Zhipu AI and MiniMax. Listed companies provide an additional, indirect signal because they also hold equity interests in private laboratories: Amazon has
invested in both Anthropic and OpenAI, Alphabet in Anthropic, and Microsoft in OpenAI. The absence of comparably large declines among these firms is therefore
informative, but not decisive. Their AI-lab holdings represent only one part of much larger
businesses, and their cloud infrastructure, distribution, product portfolios, and customer relationships may offset losses at the model layer. 

Overall, the
cross-section of returns is consistent with the paper's central thesis:
exposure depends less on whether a firm is Chinese or publishes open
weights than on its position in the value chain and its ownership of
complementary assets. Given the limitations of the event windows, this
remains an interpretation of the observed pattern rather than a causal
finding.

\section{Who Is Most Exposed to Chinese
Competition?}\label{who-is-most-exposed-to-chinese-competition}

This section provides an analysis of competitive advantages that helps to interpret the market reactions documented in the previous section. The main thesis proposed here is simple: what users purchase is not a
set of weights but a service, and the weights are only one of its
inputs. US incumbents hold advantages that do not depend on possessing
the single best model, and none of these advantages is transferred when
weights are published. Firms that occupy the \emph{same} position in the
value chain as Moonshot---providers of near-frontier models sold at
near-frontier prices, without a comparable bundle of complementary
assets (described in the following subsections)---compete on precisely the dimension on which Kimi K3 delivered,
and are therefore more directly exposed to its arrival.

\subsection{Bundling}\label{bundling}

A language model takes some text as input and predicts which words or tokens are most likely to come next. What it has learned is stored in its weights, but the quality of the final user experience also depends on how the model is prompted, connected to information, deployed, and built into a practical system.

A deployed service consists of the model \emph{plus}:

\begin{enumerate}
\def\labelenumi{\arabic{enumi}.}
\tightlist
\item
  \emph{Harness:} the orchestration layer that constructs prompts,
  manages the context window, invokes the model repeatedly, parses its
  outputs, calls tools, and handles failures. Often the harness takes the form of a \emph{task-specific
product}, for example, a coding agent such as Claude Code or OpenAI
Codex.\footnote{In these products the harness, the permissions model, and
the integration with repositories and build systems account for much of
the observed quality, and subscription packaging rather than raw token
sales sets the commercial terms.}
\item
  \emph{Tools:} external capabilities that the model can invoke, such as
  code-execution environments, web browsers, databases, and enterprise
  applications.
\item
  \emph{Grounding infrastructure:} retrieval and search systems that
  supply relevant, current, or proprietary information against which the
  model's outputs can be anchored.\footnote{A model encodes what was in
    its training corpus at the time the training run ended. In this
    respect, it is a fixed and dated object. Any task whose answer
    depends on information generated since---a price, a filing, an
    amended regulation, or the current state of a code
    repository---cannot be performed by the model alone, however capable
    it is. The relevant material needs to be located and placed in the
    model's input, which presupposes a crawled and indexed corpus, a
    ranking function to select from it, and serving infrastructure able
    to query both at a latency compatible with an interactive session.
    None of this is conveyed by publishing model weights.}
\item
  \emph{Serving infrastructure:} the accelerators, memory, networking,
  and scheduling systems required to deliver responses at an acceptable
  speed, cost, and level of reliability.
\item
  \emph{Operational and contractual guarantees:} commitments concerning
  uptime, security, data handling, regulatory compliance, and support.
\end{enumerate}

Publishing the weights therefore makes the model itself available, but
transfers none of the complementary infrastructure or assurances needed
to deliver it as a reliable service. Since users typically buy the assembled system---through
a subscription to a chat assistant, a coding agent, or an enterprise
platform---rather than raw token generation, an open-weight release of
frontier quality is a partial, not a complete, substitute for the
incumbent product. US laboratories have increasingly sold systems rather
than models, and it is the system that carries the greater margin.

\subsection{Switching Costs}\label{switching-costs}

Even where an open-weight model is a technical substitute, moving to it
is costly. US AI labs have acquired, as a first-mover advantage, vast user bases already subject to significant switching costs, which include:

\begin{enumerate}
\def\labelenumi{\arabic{enumi}.}
\tightlist
\item
  \emph{Chat and context history:} accumulated conversation history and
  the personalization built on it do not easily travel between providers.
\item
  \emph{Setup costs for specialized tools:} workflow integrations for
  agentic tools---such as Claude Code or OpenAI Codex---are
  configured, permissioned, and embedded in continuous-integration
  pipelines.
\item
  \emph{Technical configuration:} setup for Model Context Protocol (MCP)
  infrastructure, together with the credentials and access policies
  attached to them.
\item
  \emph{Ecosystem links:} integrations with third-party applications,
  prompt libraries, and internal evaluation suites tuned to specific
  model behaviors.
\end{enumerate}

The economic relevance of these switching costs is visible in
revealed-preference data. According to a recent study \cite{eastwood2026} that analyzed data gathered between May and September 2025, closed models accounted for
roughly 80\% of all tokens processed on OpenRouter, a marketplace that
routes requests across providers and where technical substitution is
relatively frictionless; yet closed models costed on average around six
times as much as open ones (\$1.86 against \$0.23 per million tokens),
and open models reached some 90\% of closed-model performance.
Reallocating usage towards open models was
estimated to be worth on the order of \$25 billion a year to the global
AI economy.

Users are, in other words, paying a substantial premium
for something other than raw capability per token: reliability, integration, and
the avoidance of migration costs. US laboratories enter each new release with
very large installed bases already subject to these costs. 

This does not however mean that the switching costs are insurmountable. When the cost differential is
sufficiently large, even high-growth firms may undertake the migration:
for instance, the startup Lindy reportedly moved 100\% of its traffic
from Claude to DeepSeek to capture savings of several million dollars
\cite{cnbc2026}. Moreover, the more recent usage data reported in
Section~\ref{ai-industry-developments-after-the-deepseek-release} partly
reverse the pattern documented by \citet{eastwood2026}: by July 2026,
closed models from the four leading US laboratories accounted for only
23\% of OpenRouter traffic, even though their absolute usage continued
to rise. Although the two measures are not directly comparable, this
shift indicates that user stickiness is not permanent. Switching costs
may slow the reallocation of workloads, but they are not insurmountable,
particularly when price--performance differentials become sufficiently
large.

\subsection{Differentiation}\label{differentiation}

Another line of defence is to avoid competing on an undifferentiated
capability metric. Four forms of differentiation are
particularly relevant.

First, laboratories often ship \emph{portfolios} rather than
single models: distinct products at different points in the
capability--speed--price space, of which Anthropic's Fable-Opus-Sonnet-Haiku and OpenAI's Sol-Terra-Luna families are examples. A rival matching one point in that space does
not automatically displace the range. Users who value speed over
reasoning depth, or cost over either, remain on a different product even
when a competitor matches the flagship.

Second, as previously discussed, models are often bundled with other services (harness, tools, grounding and serving infrastructure, guarantees). Differentiation of each of these \emph{complementary services} reinforces the competitive position of the models.

Third, laboratories differentiate on \emph{input and output modalities},
including text, images, video, speech, and documents. Models support
different subsets and combinations of these formats: one may accept
images but generate only text, while another may analyse and produce
audio or video. A model's modality profile therefore determines the
workflows it can serve and provides a further dimension of competition
that is not captured by text-based capability benchmarks.

Fourth, models differ in \emph{style and behaviour} even when their underlying
capabilities are similar. Some queries have a unique correct answer, but
many tasks---such as writing, summarisation, brainstorming, and customer
interaction---admit several valid outputs, and the preferred one depends
on the user and the context. A terse and highly technical style may suit
programmers or researchers, whereas a more conversational and accessible
style may be preferable in education, support, or consumer applications.
Models can also differ in their degree of initiative, caution,
explanation, and willingness to make assumptions, as well as in the trade-off between creativity and hallucinations. Further differences may regard the level of alignment (conformity with human intentions and values \cite{weller2026}) and censorship (e.g., in the case of Chinese models \cite{ablove2026}). These characteristics
allow providers to serve distinct preferences and use cases, making
models with similar benchmark scores imperfect substitutes.

\subsection{Trust}\label{trust}

For a significant class of customers, the binding constraint is not
capability but assurance. Financial institutions, healthcare providers,
legal services, and public administrations require documented data
handling, auditability, data residency, indemnification, and a
counterparty that can be held contractually liable. A model weight file
downloaded from a repository does not constitute a procurement-compliant
counterparty or provide legal indemnification.

Concerns of this kind are among the reasons users give for not moving to
open models \cite{eastwood2026}, and they weigh more heavily still when
a provider is subject to foreign statutory data-access mandates---such
as China's National Intelligence Law (2017), Data Security Law (2021),
and Cybersecurity Law (2017)---which mandate cooperation with state
intelligence regardless of corporate structure or server location
\cite{techtimes2026}.

Operational and security risks further compound these issues. Moonshot
AI acknowledged that Kimi K3 can display excessive proactiveness when
handling ambiguous prompts, making unexpected decisions on the user's
behalf \cite{moonshot2026}. Security audits also discovered that a previous version of Kimi had
generated privacy
failures, such as a 2026 case where Kimi disclosed a user's resume to
an unrelated party. Cases like this underscore the difficulty of providing the level of
trust required for enterprise adoption.

\subsection{Private Information}\label{private-information}

Beyond infrastructure and trust, US incumbents benefit from a ``data
moat'' consisting of proprietary information that is not publicly
available for training. The relevant assets are heterogeneous: Google
holds the web index and YouTube; xAI holds the social graph of X and,
following its recent acquisition activity, developer interaction data
from Cursor; Anthropic holds extensive usage data from enterprise
deployments; OpenAI holds the largest consumer subscription base, which
generates the specific human preference data used in reinforcement
learning from human feedback (RLHF); Meta holds its social platforms. Comparable data moats are held by Chinese conglomerates Alibaba and Tencent, which develop the Qwen and Hy (Hunyuan) models.

Equally important is the history of customer conversations, corrections,
and revealed satisfaction that every provider captures. These data are
generated as a by-product of serving and cannot be purchased or easily
replicated.

How defensible is this? The answer depends largely on whether
\emph{distillation} works: whether a competitor can recover much of the
value of a frontier model by querying it and training on its outputs.
Singh \cite{iiss2026} argues that distillation attacks may allow Chinese
laboratories to reproduce benchmark performance they have not
independently achieved, and notes that the measured gap between closed
frontier models and open-weight alternatives has narrowed.

The allegations are specific. In February 2026, Anthropic publicly
accused Moonshot AI, DeepSeek, and MiniMax of industrial-scale
extraction from its Claude models, involving more than 24,000 fraudulent
accounts and some 16 million exchanges, of which roughly 3.4 million
queries were attributed to Moonshot AI and were directed at agentic
reasoning, tool use, and coding---the same capabilities on which Kimi K3
subsequently reported its strongest results \cite{cnbcdistillation2026}.
Moonshot AI has neither confirmed nor denied that data of this kind
entered the training of K3. The claims are contested and unadjudicated.
What matters for the valuation question is the structure of the
argument: the worth of proprietary interaction data depends on the cost
of extracting it, which is a technical and legal question rather than a
property of the model itself.

Two considerations suggest that the intuition ``query the teacher, copy
the teacher'' is nonetheless weaker than it appears. First, frontier models are now trained substantially by reinforcement learning (RL),
and copying the observable behaviour of a trained RL-policy--- so-called imitation learning,
or behavioural cloning---is known to degrade badly over long tasks. The reason
is a mismatch of distributions. The imitator is trained on states the teacher
visits, but at deployment it visits the states its own mistakes lead it to,
where it has no supervision. Errors therefore compound rather than average out:
if a small per-step probability $\varepsilon$ of departing from the teacher's
trajectory is independent across $T$ steps, the probability that a task of
length $T$ contains at least one departure is

\[
1 - (1-\varepsilon)^{T} \approx \varepsilon T
\quad \text{for small } \varepsilon T,
\]

and once the trajectory has left the supervised region there is nothing to pull
it back. A model trained with RL, by contrast, has been
optimised against a reward signal on the states it actually reaches, and so has
learned how to recover from its own errors. Imitation transfers the teacher's
answers; it does not transfer the teacher's recovery behaviour, which is what
long agentic tasks require \cite{ross2011}.

Second, what is available to be imitated is shrinking. Reasoning models
generate long internal chains of intermediate steps, and it is precisely
these traces, rather than final answers, that would constitute the most
valuable training signal. Providers can and do withhold or summarise
them, exposing the conclusion while obfuscating the derivation. That
raises the cost of distillation without raising the cost of ordinary
use. In addition, frontier labs increasingly employ active
countermeasures, including platform throttling upon detecting automated
extraction and disruption of proxy networks used for systematic
distillation \cite{iiss2026}. 

Neither consideration is decisive. Benchmark scores can be approached by
distillation even where recovery behaviour is not, and benchmarks are
what markets observe. The honest conclusion is that the durability of
the data moat is an open empirical question, on which the available
evidence is still too recent to settle.

\section{A Historical Precedent}\label{a-historical-precedent}

The present competitive situation has a historical precedent close
enough to be instructive. The
diffusion of open-source software began in the early 1990s
and became a mainstream commercial phenomenon by the decade's end \cite{wikipediafoss}. The
initial reaction of incumbent vendors was to treat it as an existential
threat: if a quality substitute were available at zero cost, profits
would vanish. Three decades later, both types of software coexist profitably. Four
mechanisms account for this outcome.

First, open-source and closed-source software often catered to different
user bases and use cases. Open source was valuable when the user needed to
inspect, modify, and control the artefact itself. This configuration was
particularly useful for organizations that needeed to audit what code did,
to patch it without waiting for a vendor, to run it on hardware of one's
choosing, to avoid per-seat licences at high volume, and to avoid
dependence on the roadmap---or the survival---of a single supplier.
These users differed fundamentally from the broader market. They were
typically found among firms with significant in-house engineering
capacity, with very high-volume and cost-sensitive deployments, or
operating under data-residency or isolation constraints.
Users without these specific requirements were often buying something else: a
turnkey system, with support, documentation, training, and a
counterparty to call when it failed. Similarly, in today's AI-model market, some users need to
fine-tune models on proprietary data or to deploy them on-premises; open-weight models satisfy both needs, but they also require accelerators,
machine-learning staff, an evaluation harness, and monitoring, so that
the population able to convert published weights into a working service
is a subset---an important but small subset---of the population that
buys AI services. For everyone else, weights are an input they cannot
use. These roles for fine-tuning and on-premises deployment create a
natural market segmentation.

Second, closed-source developers could build defensible moats around
their software by integrating it with other services. Rather than
compete on license price, they moved into infrastructure, managed
databases, enhanced security, identity-management, consulting
and specialised tooling; moreover, they invested in retention, contractual
relationships, and product differentiation in order to avoid selling an
undifferentiated good. The five components that surround an AI model in a
deployed service (Section 4.1) are the same manoeuvre applied to a new
commodity.

Third, demand grew so much that it created opportunities for everybody.
Cheap components lowered the cost of building products, raising the
quantity demanded of the whole stack. For example, free web-server software helped make the
commercial internet possible, which in turn generated demand for
proprietary databases, middleware, and cloud computing. Growth in demand
for the overall technology stack more than offset the loss of value in
the layer that was commoditised.

Fourth, closed-source developers often embraced open source and
transformed it into a profit opportunity. Microsoft moved from open
hostility to shipping and supporting Linux on its cloud platform
(Azure); Amazon Web Services built a business around open engines it did
not own; Oracle, whose main product was a database management system, acquired the developer of MySQL, one of the most popular open-source alternatives to its system. The general principle is that a firm
which controls a bottleneck---distribution, infrastructure, or the
customer relationship---benefits from the commoditisation of everything
adjacent to it. Cheaper complementary inputs reduce the cost of the
service and expand its use, while the bottleneck owner can continue to
charge for access and capture part of the value created. The release of open-weight models by US laboratories,
discussed in Section 6, is the same move.

Crucially, this historical pattern suggests that competition between
open- and closed-source software fosters a healthy ecosystem in which
innovation does not stall. The pressure from open alternatives served as
a constant stimulus for closed-source developers to innovate and
differentiate, rather than as a precursor to industry stagnation.

Two qualifications are in order. First, the analogy is imperfect,
because open weights are not a volunteer commons: the releases discussed
here are corporate products of capital-intensive training runs,
published for strategic reasons. Second, commoditisation was not
costless for the firms whose core product was displaced: it
redistributed rents across the value chain rather than preserving them
at the software layer.

\section{AI Industry Developments after the Two
Releases}\label{ai-industry-developments-after-the-two-releases}

What has happened since the two releases is broadly consistent with the
mechanisms set out above, and it suggests that several of the
competitive advantages described in Section 4 are solidly in place.

\subsection{Developments after the release of DeepSeek R1}\label{ai-industry-developments-after-the-deepseek-release}

\emph{Open-weight usage rose sharply---and so did closed-model usage.}
Table \ref{tbl-openrouter-token-usage} reports estimates based on
OpenRouter data, a marketplace that routes API requests across model
providers.

\begin{table}[H]
\centering
\caption{Estimated weekly token usage on OpenRouter}
\label{tbl-openrouter-token-usage}
\vspace{0.3em}
\small
\renewcommand{\arraystretch}{1.25}
\begin{tabular}{@{}>{\raggedright\arraybackslash}p{0.22\textwidth}
                    >{\centering\arraybackslash}p{0.20\textwidth}
                    >{\centering\arraybackslash}p{0.30\textwidth}
                    >{\centering\arraybackslash}p{0.18\textwidth}@{}}
\toprule
\textbf{Week beginning} &
\shortstack{\textbf{Total usage}\\\textbf{(trillion/week)}} &
\shortstack{\textbf{Top-four US labs}\\\textbf{(trillion/week)}} &
\shortstack{\textbf{Top-four}\\\textbf{share}} \\
\midrule
20 Jan. 2025 & 0.53 & 0.34 & 64\% \\
19 Jan. 2026 & 7.49 & 4.61 & 62\% \\
20 Jul. 2026 & 58.00 & 13.19 & 23\% \\
\bottomrule
\end{tabular}
\par\vspace{0.5em}
\parbox{\dimexpr0.90\textwidth+6\tabcolsep\relax}{\footnotesize\textit{Note:} Estimates based on OpenRouter data. The top-four US laboratories considered here are Anthropic, Google, OpenAI and xAI. Open models released by these laboratories are excluded from the statistics on top-four labs. The market share of closed models by the top-four labs is calculated relative to total weekly token usage.}
\end{table}

Total weekly usage rose from 0.53 trillion tokens per week in January 2025 to
58 trillion in July 2026. Usage of closed models released by the top US
laboratories also increased sharply, from 0.34 trillion to 13.2 trillion
tokens per week. Their absolute usage therefore grew nearly forty
times, even as their share of total traffic fell from 64\% to 23\%.
Open and other models expanded faster and gained substantial market
share, but this occurred in a rapidly growing market rather than through
an absolute contraction in closed-model usage. These estimates
therefore indicate that usage of both open and closed models surged after
the release of DeepSeek R1.

\emph{Compute demand did not fall.} Neither release was followed by a
decline in the demand for accelerators. Cheaper tokens raised the volume
of tokens consumed, and the shift towards reasoning and agentic
workloads raised the number of tokens required per task. The revenues of NVIDIA, the primary supplier of AI accelerators, grew from 44 USD billion in the quarter ending in April 2025 to 82 billion in the same quarter of 2026. This is 
Jevons paradox in practice: efficiency gains in the use of a resource
often increase rather than decrease its total consumption. 

\emph{US companies released open-weight models of their own.} OpenAI released the
GPT-OSS family of open-weight reasoning models, very fast models designed to run
on minimal infrastructure, including consumer GPUs \cite{openai2025gptoss}. Google DeepMind released the Gemma family
of lightweight open-weight models, built from the
same research and technology used for their flagship closed models and intended for deployment
on hardware ranging from mobile devices to cloud servers
\cite{google2026gemma}. NVIDIA disclosed the Nemotron family, more explicitly
oriented towards developers and enterprises building specialised AI
agents; NVIDIA published not only weights but, for some models, training
data and recipes \cite{nvidia2025nemotron}. These releases illustrate a strategic behavior similar to that observed in the software industry in the Nineties, in response to the diffusion of open-source software: incumbents
can supply the open layer themselves while continuing to monetise the
infrastructure, tools, and customer relationships surrounding it.
Chinese open-weight competition likewise broadened beyond DeepSeek after
the release of R1. Subsequent examples included Alibaba's Qwen3,
Moonshot AI's Kimi K2, Z.ai's GLM-4.5, and MiniMax's M2
\cite{qwen2025qwen3,moonshot2025k2,zai2025glm45,minimax2025m2}. This
proliferation is relevant context, although the main point here is the
decision by US incumbents to participate directly in the open-weight
segment.

\subsection{Developments after the release of Kimi K3}\label{ai-industry-developments-after-the-kimi-release}

\emph{Capability without capacity is not a complete product.} Within
days of the Kimi K3 release, Moonshot AI temporarily paused new
subscriptions after demand exceeded its available serving capacity
\cite{kimipause}. This is the bundling argument in a particularly
direct form: the fourth component of a deployed service---serving
infrastructure---became binding. Frontier-class weights retained
economic value, but the provider could not extend access to all
prospective users without sufficient compute and operational capacity.

\emph{Competitive responses included rapid new model releases.} At the
time of its release, Kimi K3 had a higher Artificial Analysis
Intelligence Index score than Anthropic's Claude Opus 4.5. A few days
later, Anthropic released Claude Opus 5, which overtook Kimi K3 and moved
to the top of the leaderboard. Google also introduced new models,
including Gemini 3.6 Flash, which combined high capability with unusual
speed and efficiency by frontier-model standards. Some commentators have
speculated that US laboratories maintain a reserve of trained models
that can be released in response to a competitor's advance. Although
the observed sequence is consistent with such a strategy, no public
evidence confirms it. It nevertheless illustrates the ability of US
laboratories to respond rapidly to competitive challenges at the
capability frontier.

\subsection{Remarks about industry developments}\label{ai-industry-developments-after-the-releases-remarks}

None of this implies that the competitive threat is illusory. The gap
between the open-weight and closed frontiers remains narrow, and open
models already match closed-model performance on several economically
valuable tasks at a small fraction of the cost \cite{clark2026}. Nor is
there a visible tendency for the closed frontier to pull away
\cite{clark2026}. It cannot be excluded that future open-weight releases could overtake
the closed frontier while retaining a substantial price advantage.
There is also evidence that US firms are shifting workloads to Chinese
models in order to capture significant cost savings \cite{cnbc2026}.
The scale of the difference for an individual user is illustrated by
one developer's estimate that an hour of assisted coding costs about
\$10 on a leading US model, compared with less than 50 cents on DeepSeek
\cite{macgpu2026}. The evidence in this paper therefore supports a claim
about \emph{where} competitive pressure currently falls---primarily on
the model layer within a broader bundle of services---not a claim that
such pressure is absent. Nor are US laboratories guaranteed to be
immune. A firm whose revenues or profits depend disproportionately on
the commoditised model layer may experience margin erosion before it can
build sufficiently strong moats around orchestration, distribution,
proprietary data, or the customer relationship. Moreover, the layers
that remain differentiated today need not remain so permanently.
Serving infrastructure, for example, could itself come under greater
pressure if Chinese accelerator producers became credible competitors
at scale. In other words, competitive pressure could migrate across the stack
as technologies mature and new suppliers emerge.

\section{Conclusions}\label{conclusions}

Two releases of Chinese language models, eighteen months apart,
coincided with headline equity movements of almost identical size and
very different composition. The January 2025 episode was, in substance,
news about the production function for artificial intelligence: if
frontier-class reasoning could be trained and served at a small fraction
of the assumed outlays, then the projected demand for accelerators---and
the capital expenditure predicated on it---was too high, and the compute
supply chain was repriced accordingly. The July 2026 episode was news of
a different kind. Kimi K3 achieved frontier-level capability and was
priced accordingly; once token consumption and output speed are
considered, it offered no clear economic advantage over
competing frontier models. What the release
revealed was therefore not a cheaper technology but an additional
credible supplier at an existing price point. After the event, the largest equity-price declines
were reported by other Chinese
laboratories. 

The analysis in this paper suggests why that cross-section is not
anomalous. What is sold is a service, of which the weights are one input
among several: harness, tools, grounding, serving capacity, and
contractual assurance are not transferred when weights are published.
Switching costs are substantial, as revealed by the persistent
willingness of users to pay several times the open-model token price;
differentiation and portfolio strategies move competition off the single
dimension that an open release commoditises; and the segment of demand
that requires an accountable counterparty is, for now, poorly served by
downloadable artefacts. 

Industry developments after the release of DeepSeek R1 are consistent with this
reading: open-weight usage and closed-model usage both grew, compute
demand did not fall, US laboratories published open-weight models of
their own. 

The closest historical precedent, the
diffusion of open-source software from the early 1990s, points in the
same direction: the arrival of a free substitute for one layer of a
stack redistributes rents across the stack rather than eliminating them,
and incumbents that control distribution, infrastructure, or the
customer relationship have generally ended up monetising the free layer
rather than being displaced by it.

For those pricing the AI complex, the practical implication is that the
relevant state variable is not the ranking of models on a public
leaderboard, which is revised every few weeks, but the ownership of the
complements that a leaderboard does not measure. This conclusion does not justify complacency. First, the open-weight frontier remains close
to the closed frontier, open models already match closed models on several
economically valuable tasks at much lower prices, and some US firms have
shifted workloads to Chinese models to capture these savings. It cannot be ruled out that future
open releases could surpass the closed frontier without
abandoning their price advantage. Second, the competitive protection enjoyed by US labs is not guaranteed. Those whose revenues
depend too heavily on access to the model layer may suffer margin erosion
before they can establish strong positions in orchestration,
distribution, proprietary data, serving infrastructure, or customer
relationships. Nor are these complementary layers permanently protected:
serving infrastructure, for example, could itself become more
competitive if Chinese accelerator producers achieved sufficient scale
and performance. 

The analysis in this paper has focused on market mechanisms, but the
competitive landscape is also shaped by industrial policy and national
security. Chinese AI laboratories operate within an ecosystem supported
by central and local government measures, including state-guided equity
funds, public investment in computing infrastructure, and vouchers that
reduce eligible firms' costs of accessing computing power, data corpora,
and model services \citep{rand2025chinaai,shanghai2025aivouchers}. Such support can
lower the cost of capital and accelerate entry, scaling, and diffusion.
It should not, however, be treated as evidence that the reported training
cost or API price of any particular model is simply a subsidy: engineering
efficiency, commercial strategy, and public support are distinct channels
that are difficult to separate with the available information.

Western governments have simultaneously debated restricting Chinese AI
models on privacy, cybersecurity, censorship, and strategic-dependence
grounds. Australia prohibited DeepSeek products on federal-government
systems in February 2025, and the Czech government adopted a comparable
restriction for state administration in July 2025
\citep{australia2025deepseek,czech2025deepseek}. In the United States, the
bipartisan No Adversarial AI Act proposed barring federal agencies from
acquiring or using AI developed by foreign-adversary entities, subject to
specified exceptions \citep{uscongress2025adversarialai}. As of August 2026, however, Congress had not passed
the proposal into law; it therefore imposed no restriction on federal
agencies and no general prohibition on private use.
The policy trade-off is broader than a contest between national
champions. On the one hand, restrictions can reduce the exposure of sensitive data and
public institutions to providers subject to foreign-state influence. On the other hand, 
blanket prohibitions would also deny Western firms access to low-cost
models, weaken competitive pressure on domestic suppliers, and potentially
slow diffusion. In July 2026, a vast coalition of US tech firms, including most US AI labs, signed an open letter highlighting exactly these drawbacks \cite{openweights2026}. Moreover, open weights complicate provider-based rules:
an open-weight model can be run on trusted domestic infrastructure without transmitting
prompts to its original developer, although risks embedded in the weights,
training choices, and downstream software still require independent
evaluation. A proportionate policy should consequently distinguish remote
services from locally hosted weights, ordinary commercial workloads from
sensitive uses, and verifiable technical risks from protection of domestic
producers.

\end{document}